\documentclass[10pt,aps,prx,raggedbottom,longbibliography,reprint,citeautoscript,letterpaper]{revtex4-2} 
\usepackage[usenames,dvipsnames]{color}
\usepackage{graphicx,microtype}
\usepackage[bookmarks=false,colorlinks]{hyperref}
\hypersetup{linkcolor=magenta,citecolor=MidnightBlue,filecolor=Plum,urlcolor=MidnightBlue}
\usepackage[all]{hypcap} 
\usepackage{lmodern}
\usepackage{amsmath}
\usepackage{multirow}
\usepackage{soul}
\usepackage{comment}



\begin{document}
\title{Lattice-Renormalized Tunneling Models for Superconducting Qubit Materials}

\author{P.\ G.\ Pritchard}
\affiliation{Department of Materials Science and Engineering, Northwestern University, Evanston, IL, USA}

\author{James M.\ Rondinelli}
\email{jrondinelli@northwestern.edu}
\affiliation{Department of Materials Science and Engineering, Northwestern University, Evanston, IL, USA}

\date{\today}

\begin{abstract}
We present a lattice-renormalized formalism for configurational tunneling two-level systems (TLS) that overcomes limitations of minimum-energy-path and light-particle models.
Derived from the nuclear Hamiltonian, our formulation introduces composite phonon coordinates to capture lattice distortions between degenerate potential wells. 
This approach resolves deficiencies in prior models and enables accurate computation of tunnel splittings and excitation spectra for hydrogen-based TLS in bcc Nb. 
Our results bound experimental tunnel splittings and reveal strong anharmonic couplings between tunneling atoms and lattice phonons, establishing a direct link between TLS dynamics and phonon-mediated strain interactions. 
The formalism further generalizes to multi-level systems (MLS), providing insight into defect-induced decoherence in superconducting qubits and guiding strategies for materials design to suppress TLS-related loss.
\end{abstract}

\maketitle



\section{Introduction}

The coherence of superconducting qubits and related quantum devices is fundamentally limited by parasitic interactions between the qubit state and its surrounding environment. 
Extensive experimental and theoretical studies have established that such couplings originate from materials defects in substrates, superconducting films, and surface oxides \cite{Zhang2024, Virginia2022, Martinis2005, Mller2019, Crowley2023}.
These findings highlight the critical roles of materials selection, growth conditions, and processing strategies on reducing defect densities for improving device performance \cite{Bal2024, McFadden2025, Place2021, Garcia-Wetten2025, Pritchard2025}.
A particularly important class of defects are configurational two-level systems (TLS), which arise from nearly degenerate atomic configurations separated by small energy barriers \cite{Mller2019}. 
These defects exhibit energy splittings comparable to or smaller than typical qubit transition energies, enabling both coherent \cite{Lisenfeld2015, Liu2024} and incoherent interactions with the qubit state \cite{deGraaf2021, Mller2019}. 
Their presence introduces dissipation and noise that degrade qubit fidelity, making accurate modeling of TLS energies essential for mitigating decoherence.

The impact of a configurational TLS on qubit performance depends on whether its transition frequency lies near the operating frequency of the qubit. 
Two approaches are commonly used to estimate such frequencies: ($i$) the minimum-energy-path (MEP) method and ($ii$) the light-particle approximation. 
The first employs the nudged elastic band (NEB) algorithm to compute the MEP between two degenerate configurations and then solves a one-dimensional (1D) Schr\"odinger equation along this path \cite{Holder2013, Wang2025}. 
The second treats the tunneling atom as a light particle embedded in a rigid lattice, computing its 3D potential energy surface for a symmetrized structure and solving the corresponding Schrödinger equation \cite{Gordon2014}. 
While both methods provide approximate tunnel splittings, they neglect essential features of the nuclear Hamiltonian. For instance, as we show, the light-particle formalism relies on thermodynamically unstable atomic configurations \cite{Gordon2014}.

The aforementioned limitations are well documented.
Ring-polymer calculations show that the MEP is not necessarily the most efficient tunneling path, implying that tunnel splittings derived from it should systematically underestimate transition frequencies \cite{Nandi2023}. 
Moreover, a 1D Schr\"odinger equation with a fixed effective mass is ill-defined along a path with continuously varying nuclear coordinates. 
Similarly, the light-particle approach assumes symmetrized structures that are energetically unrealistic: for example, the symmetrized structures used to model hydrogen TLS in bcc Nb exhibit formation energies exceeding 10\,meV 
($E_f=E_c>0$, see below); values far above thermal energy available in quantum circuits operating below 100\,mK. 
Even larger formation energies ($>\!30$\,meV) were reported for symmetrized structures in Al$_2$O$_3$ \cite{Gordon2014}. 
These shortcomings motivate the need for a tunneling model that rigorously incorporates lattice degrees of freedom while preserving thermodynamic stability.

Here we introduce a formalism derived directly from the nuclear Hamiltonian that overcomes limitations of existing minimum-energy-path and light-particle approaches. 
Our method incorporates composite phonon coordinates to capture lattice distortions between degenerate wells, enabling a rigorous treatment of tunneling dynamics.
Applied to hydrogen TLS in bcc Nb, the formalism produces tunnel splittings that bound experimental transition frequencies from below and above and reveals a fundamental connection between configurational TLS and lattice phonons beyond conventional strain coupling.
It further generalizes to H multi-level systems arising from clusters of three or more spatially-localized, degenerate, configurations. 
We conclude by discussing the implications of our results for defect mitigation in superconducting qubit applications.

\begin{figure}
\centering
   \includegraphics[width=0.35\textwidth]{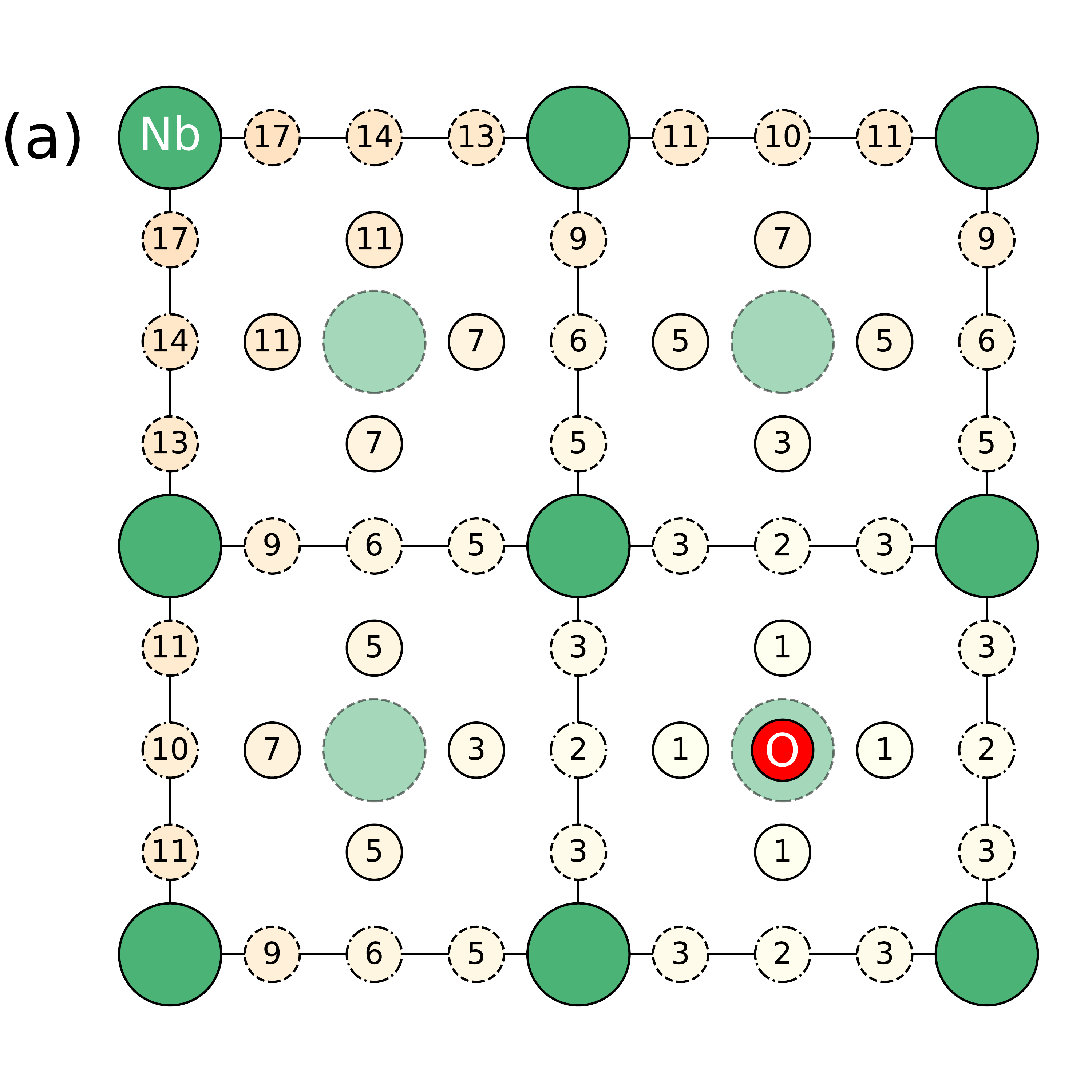}
   \includegraphics[width=0.35\textwidth]{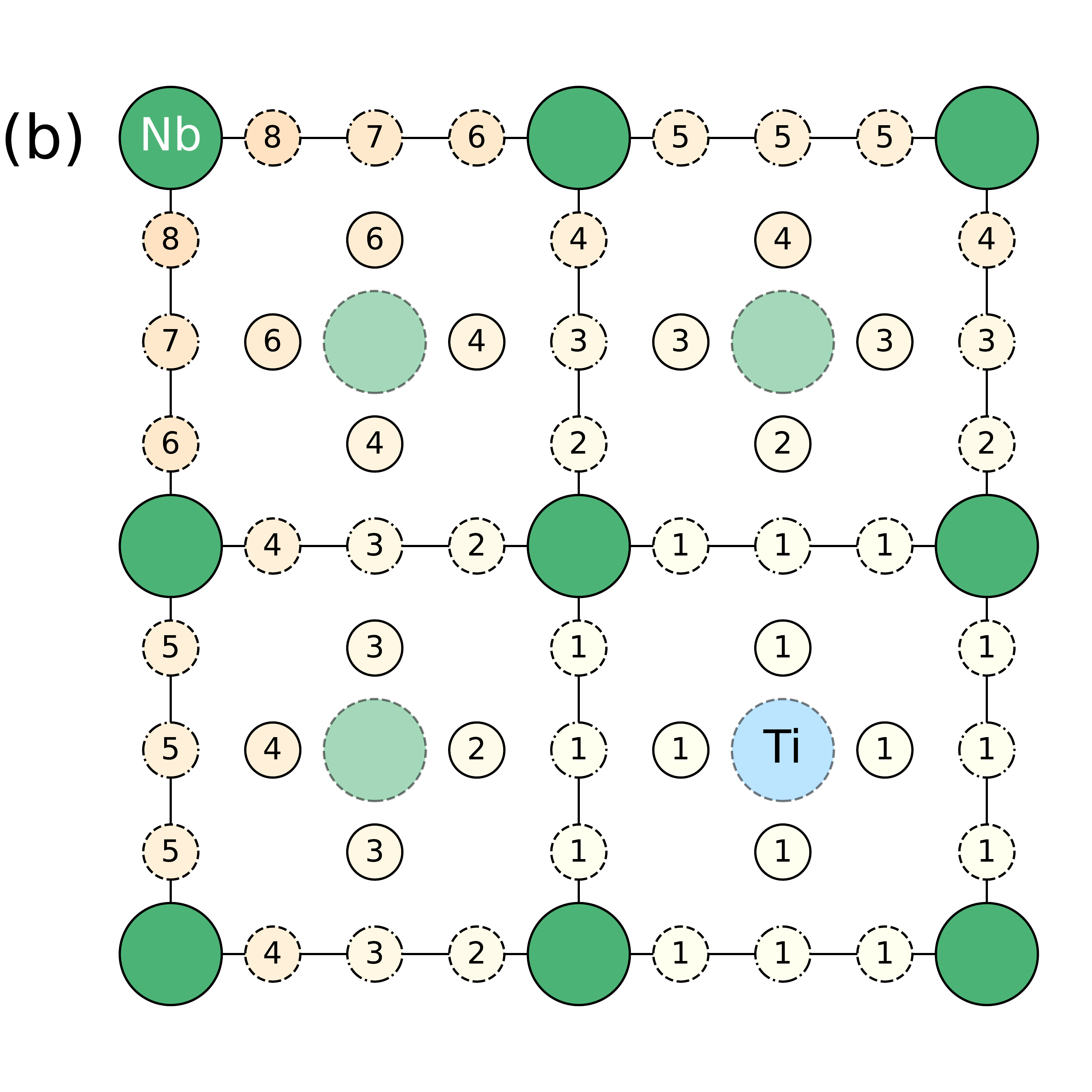}
   \caption{(a, b) H interstitial sites in BCC Nb within $2\times2\times1/2$ supercell. The out-of-plane position is indicated by solid ($z=0$), dash-dotted ($z=a/4$), and dashed lines ($a/2$). H sites are labeled and shaded by their distance from an (a) interstitial O site or a (b)  substitutional Ti site. }
   \label{fig:h_sites}
\end{figure}


\section{Materials System and Defect Sites}
A prior work has demonstrated that interstitial, self-trapped H favors tetrahedral sites in bcc Nb \cite{Sundell2004}; however, at low temperatures, self-trapped H precipitates into ordered niobium hydrides \cite{Magerl1983, Leibengood2025, TorresCastanedo2024, Garcia-Wetten2025}. 
Consequently, most H atoms that exhibit tunneling behavior must be trapped by other point defects such as O, Ti, and Zr \cite{Neumaier1982,Wipf1984,Magerl1986,Cordero2022}.
We therefore apply our formalism to self-trapped H and H trapped by adjacent O, Ti, and Zr. Oxygen, which occupies interstitial octahedral sites in the bcc structure \cite{Reynolds2024}, is well-known to exhibit hydrogen tunneling behavior \cite{Wipf1984, Magerl1986}, whereas Ti and Zr are substitutional defects that also exhibit hydrogen tunneling behavior \cite{Neumaier1982, Cannelli1994,Cordero2022}. We also consider the potential for H to be trapped by substitutional Ta whose presence in Nb films may be expected due to its chemical similarity to Nb and possible role as a encapsulation layer for Nb films \cite{Bal2024}.

\begin{table}[]
    \caption{Formation energy ($E_f$) of an O-H complex in bcc Nb with H located at the 1\textsuperscript{st} to 8\textsuperscript{th} nearest-neighbor sites relative to the O site ($c=1/128$). $N_d$ is the number of adjacent, degenerate H sites in the nearest neighbor shell, $r_z$ is the component of the $\mathbf{r}$ vector connecting O to H parallel to the axis of expansion of the O defect, $r$ is the magnitude of $\mathbf{r}$.}
    \begin{ruledtabular}
    \begin{tabular}{clccc}
    Site   &$N_d$&  $r_z$& $r$& $E_f^\mathrm{O\textrm{-}H}$\\
     -   && (\AA) & (\AA) & (meV) \\ 
   \hline
    1      &4& 0.00 & 0.83 & unstable \\
    2      &1& 0.83 & 1.85 & 374      \\
    3a     &2& 1.65 & 2.48 & 118      \\
    3b     &1& 0.00 & 2.48 & 47.5     \\
    4      &1& 2.48 & 2.98 & 136      \\
    5a     &2& 1.65 & 3.41 & 5.3      \\
    5b     &1& 0.00 & 3.41 & 28.8     \\
    5c     &4& 3.31 & 3.41 & 75.8     \\
    6      &1& 0.83 & 3.79 & -20.3    \\
    7a     &2& 0.00 & 4.14 & -72.5    \\
    7b     &1& 3.31 & 4.14 & 34.2     \\
 7c& 1& 0.00& 4.14 &-9.0\\
    8a&1& 2.48& 4.45& -2.6\\
 8b& 1& 4.13& 4.45&40.3\\
 \end{tabular}
 \end{ruledtabular}    
 \label{tab:OH_ef}
\end{table}
 
\autoref{fig:h_sites} presents the distribution of H interstitial sites relative to an interstitial O site and a substitutional Ti site within a $2\times2\times1/2$ supercell of bcc Nb. Each H site is indexed by its relative distance to the defect site. To identify the preferred H interstitial positions adjacent to an O interstitial site, we computed the formation energy of OH complexes ($E_f^\mathrm{OH}$) for all H sites up to the 8th nearest neighbor H sites in a $4\times 4\times 4$ supercell as 
\begin{equation}
\label{eq:ef}
E^\mathrm{OH}_f=E_\mathrm{OH}-E_\mathrm{H}-E_\mathrm{O}+n_\mathrm{Nb}\mu_\mathrm{Nb}\,
\end{equation}
where $E_\mathrm{OH}$, $E_\mathrm{O}$, and $E_\mathrm{H}$ are the total energies of the OH complex, a single O interstitial, and a single H insterstitial computed using density functional theory (DFT). The term $n_{Nb}$ denotes the number of Nb atoms in the supercell, and the Nb chemical potential, $\mu_\mathrm{Nb}$, is referenced to the total energy per atom of bcc Nb.

\autoref{tab:OH_ef} summarizes the O-H complex  formation energies, the degeneracies of corresponding adjacent H sites, and the position of the tetrahedral sites relative to the O interstitial. We find that the favored site is a 7th nearest neighbor tetrahedral site with a  formation energy of -72.5\,meV. This value is in reasonable agreement with the experimentally  reported binding enthalpy of $\sim$100 meV \cite{Wipf1984}.  These data show that H unambiguously occupies a well-defined tetrahedral site in the presence of an O trap. At the millikelvin  operating temperatures relevant for superconducting qubits, only this site will be thermodynamically accessible, and therefore we use it when computing the tunnel splitting of O-trapped H in the remaining work. 

The preference for this site can be understood on two complementary grounds. First, the elastic dipole tensors for H and O interstitials in bcc Nb take the form: 
\[
P_\mathrm{H}=\begin{bmatrix}
A & 0 & 0 \\
0 & B & 0 \\
0 & 0 & B \\
\end{bmatrix},
\space
P_\mathrm{O}=\begin{bmatrix}
C & 0 & 0 \\
0 & -D & 0 \\
0 & 0 & -D \\
\end{bmatrix}
\]
For H, $A\approx B$ \cite{Sugimoto1980} implies an almost isotropic local expansion of the lattice. 
%
In contrast, O interstitials exhibit an axis of expansion and a perpendicular plane of contraction. 
An H atom therefore minimizes its elastic interaction energy by occupying a site within the O-induced contraction plane, consistent with the DFT results.
Second, O occupies a negatively charged (O$^{2-}$) octahedral site in bcc Nb. The $2p^6$ valence electrons of the oxide ion preferentially align with the 6 adjacent Nb sites; however, as the bcc octahedral sites are anisotropic, the Nb--O bonds distort to equalize their lengths. This results in an expansion of the bond along the short axis and compression of the bond along the long axes. The presence of H at the optimal tetrahedral sites enhances the compression of the Nb--O bond along the long axes, stabilizing the configuration relative to alternative sites.
We anticipate similar mechanisms will apply to other octahedral interstitials such as carbon and nitrogen.

\begin{table}[]
    \caption{Formation energy ($E_f$) of Ti-H, Zr-H, and Ta-H complexes in bcc Nb with H located at the 1\textsuperscript{st} to 5\textsuperscript{th} nearest-neighbor sites relative to the O site  ($c=1/54$). $N_d$ is the number of adjacent, degenerate H sites in the nearest neighbor shell, $r$ is distance between the H site and the substitutional defect.}
    \begin{ruledtabular}
    \begin{tabular}{clccll}
    Site   &$N_d$& $r$& $E_f^\mathrm{Ti\textrm{-}H}$& $E_f^\mathrm{Zr\textrm{-}H}$&$E_f^\mathrm{Ta\textrm{-}H}$\\
     -   && (\AA) & (meV)  & (meV)  &(meV)  \\ 
   \hline
    1      &24& 1.85& -73.8& -50.0&27.1\\
    2      &2& 2.98& -0.4& -39.9&10.0\\
    3&2& 3.79& 4.7& -9.8&1.6\\
    4a&6& 4.45& 8.4& 25.3&0.5\\
    4b&1& 4.45& -0.2& -1.9&-2.83\\
 5& 4& 5.01& -0.1& 33.3&-15.9\\
 \end{tabular}
 \end{ruledtabular}
    
    \label{tab:sub_ef}
\end{table}

\begin{figure*}
\centering
\includegraphics[width=0.35\textwidth]{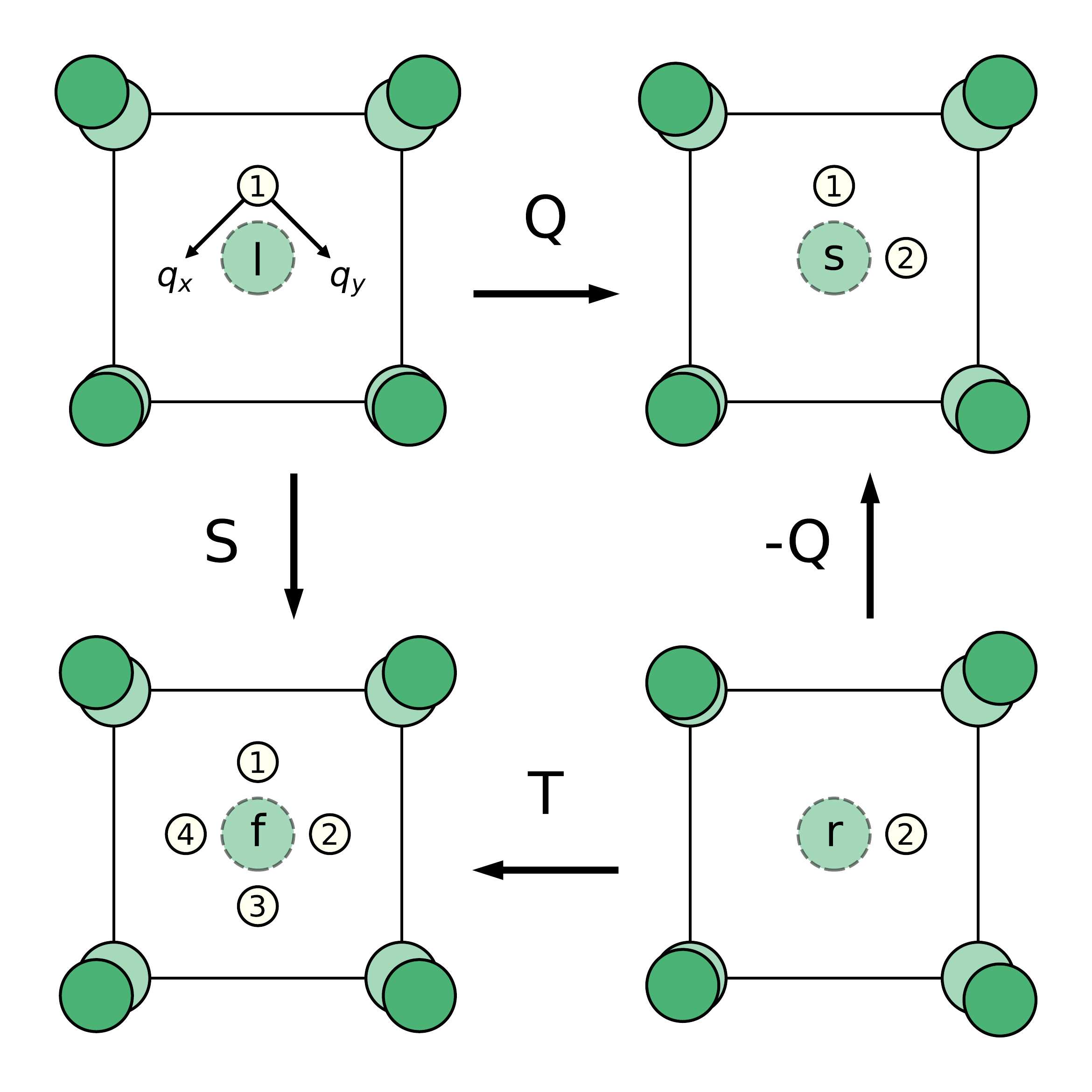}\vspace{12pt}
   \includegraphics[width=0.64\textwidth]{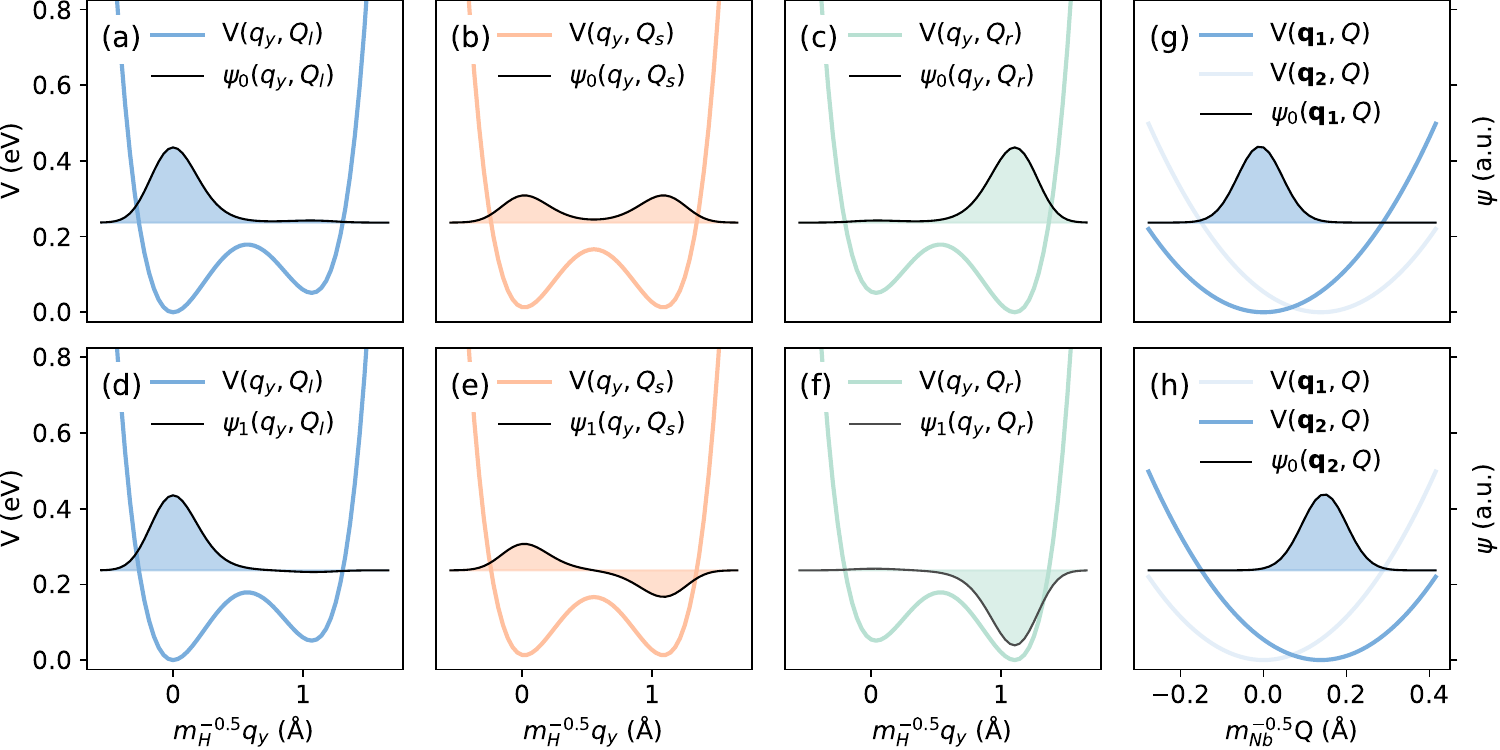}
   \vspace{-24pt}
   \caption{%
   (left) Schematic of lattice order parameters. Nb (green) and degenerate H tetrahedral sites (white) are shown; offset Nb atoms indicate lattice distortions. Atoms with solid borders lie in the same plane; the dashed Nb is body-centered. Order parameter $Q$ transforms singly-degenerate H configurations ($l$ or $r$) into a two-fold degenerate configuration ($s$); $S$ and $T$ produce a four-fold degenerate configuration ($f$).
   (a-f) Potential energy ($V$), ground ($\psi_0$) and first-excited ($\psi_1$) state wavefunctions for an O-H defect along the path connecting adjacent, degenerate H sites. $V$, $\psi_0$, and $\psi_1$ are shown at three fixed values of the composite phonon coordinate ($Q_l$, $Q_{s}$, $Q_r$). (g, h) $V$ and $\psi_0$ are shown along the composite phonon coordinate $Q$ at fixed hydrogen coordinates ($\mathbf{q_1}$, $\mathbf{q_2}$).}
   \label{fig:h_phi}
\end{figure*}

Owing to their cubic symmetry, substitutional  impurities such as Ti, Zr, and Ta, create significantly  simpler H-site environments than interstitial O (\autoref{fig:h_sites}). 
Their elastic dipole tensors are isotropic, so the long-range elastic interactions between H and a substitutional defect is governed primarily by the sign and magnitude of the defect's elastic dipole tensor rather than strain-field anisotropy. 
We computed the formation energies of H bound by substitutional Ti, Zr, and Ta defects using a $3\times3\times3$ supercell using \autoref{eq:ef} with appropriate modifications.
\autoref{tab:sub_ef} summarizes the computed formation energies 
up to the 5th nearest neighbor sites. 
For both Ti-H and Zr-H complexes, the first-nearest neighbor tetrahedral site exhibits the lowest formation energy, with Ti-H having a $E_f$ comparable to that of the O-H complex. 
Zr-trapped H is slightly less stable with $E_f=-50.0$\,meV, however its stability extends further, up to the third-nearest neighbor, in contrast to Ti-trapped H, which is essentially only stable at the first-nearest neighbor site.
In contrast, Ta defects repel H. The Ta-H formation energy remains positive through the fourth-nearest neighbor sites, indicating that H is excluded within in region of radius $\sim$5 \AA\ around Ta in Nb. 
%
%
Based on these results, we restrict our tunnel-splitting calculations to Ti-trapped and Zr-trapped H in the present work.

\section{Tunneling Model}

\subsection{Theory}
To describe tunneling between degenerate lattice configurations, we start from the Born-Oppenheimer nuclear Hamiltonian and reduce it to the essential degrees of freedom. The resulting low-dimensional Hamiltonian explicitly couples the hydrogen defect to a collective lattice mode that mediates transitions between wells:
\begin{equation}\label{eq:reduced}
    \hat H_n^{\text{sub}} = \sum_{i=\mathbf{q},Q}\frac{-\hbar^2}{2}\nabla_i'^2 + V(\mathbf{q},Q),
\end{equation}
where $\mathbf{q}$ denotes the hydrogen phonon coordinates, $Q$ is a mass-normalized composite lattice coordinate that continuously transforms the host lattice between adjacent degenerate configurations, and $\nabla_i'$ are the mass-scaled gradient operators for $(\mathbf{q},Q)$.
The nuclear potential $V(\mathbf{q},Q)$ is exact within the confines of the Born-Oppenheimer approximation and the accuracy of density-functional theory. It contains the harmonic contributions of the interstitial hydrogen and composite lattice mode and the anharmonic couplings between $\mathbf{q}$ and $Q$. 
Unlike conventional approaches, $V(\mathbf{q},Q)$ is not restricted to either the minimum-energy path or a static lattice.
Instead, it is sampled on a multidimensional grid spanning two degenerate wells, allowing us to capture the energetically relevant coupling surfaces.
Here, we ignore the cross-coupling to the remaining phonon modes $\{\omega_{iQk}\}_{i\neq \mathbf{q},Q}$ as they weakly influence the computed tunnel splitting, as we argue below.
This formulation retains the essential anharmonicity required for accurate tunneling predictions and avoids reliance on thermodynamically unstable symmetrized structures. 
Physically, the dominant anharmonicity arises from the hydrogen phonon modes and the composite coordinate $Q$, which parameterizes the Nb lattice distortion between degenerate configurations 
(\autoref{fig:h_phi}, left) and discussed further below.
The full derivation, including the transformation from the nuclear Hamiltonian and its expansion in phonon coordinates, is provided in the Appendix. 
Note that in the limit where $Q$ is fixed at its symmetric value, $Q_s$, \autoref{eq:reduced} reduces to a 3D H coordinate spanning both wells--the conventional light-particle picture in a rigid lattice. 
%

\subsection{Numerical Implementation}
To solve \autoref{eq:reduced}, we compute $V(\mathbf{q},Q)$ on a four dimensional grid spanning two degenerate H sites using density functional theory (DFT) \footnote{All electronic structure simulations were performed with the Vienna Ab initio Simulation Package (VASP) \cite{Kresse1996, Kresse1999} and the Perdew-Burke-Ernzerhof (PBE) generalized gradient approximation (GGA) functional \cite{Perdew1996}. This functional has been previously shown to accurately capture the structure and elastic properties of bcc Nb \cite{v27k-y5fb}. The projector augmented wave (PAW) pseudopotential method \cite{Blochl1994} was used to treat core and valence electrons with the following electronic configurations: 
4$s^2$4$p^6$5$s^1$4$d^4$ (Nb),
3$s^2$3$p^6$4$s^1$3$d^3$ (Ti),
4$s^2$4$p^6$5$s^1$4$d^3$ (Zr),
2s$^2$2p$^4$ (O), and 
1s$^1$ (H).
The energy cutoff was 600 eV and the specified $k$ point grids were equal to $\lceil\frac{16}{n}\rceil\times\lceil\frac{16}{n}\rceil\times\lceil\frac{16}{n}\rceil$, where $n$ is the supercell dimension. Methfessel-Paxton smearing was specified with $\sigma=0.2$ eV. Structures were relaxed to a force tolerance of 1 meV\,\AA$^{-1}$.}.
The grid is parameterized by $(\hat q_x,\hat q_y,\hat q_z,\hat Q)$, corresponding to the three H phonon modes $\mathbf{q}$, specified with Cartesian axes, and the composite lattice mode $Q$.
The origin $(0,0,0,0)$ is defined as the fully relaxed structure with H at one of its lowest-energy sites (e.g., site 1). 
We orient $\hat q_y$ along the path connecting the two degenerate H sites, $\hat q_z$ normal to the $m_z$ mirror plane, and $\hat q_x=\hat q_y\times\hat q_z$ (\autoref{fig:h_phi}).
The composite coordinate is defined as $\hat Q=\mathbf{Q}/|\mathbf{Q}|$ and is orthogonal to $(\hat q_x,\hat q_y,\hat q_z)$. 

We solve the 
Schr\"odinger equation by defining a finite difference Hamiltonian on a subgrid whose spacings are integer multiples of the sample grid's spacings such that $q_x,q_z\in[-q_{y_{12}}/2,q_{y_{12}}/2]$, $y\in[-q_{y_{12}}/2,3q_{y_{12}}/2]$, and $Q\in[-2Q_{lr},3Q_{lr}]$, sampled with $7\times13\times7\times11$ evenly spaced points.
Here, $q_{y_{12}}$ and $Q_{lr}$ denote the distances along $\hat q_y$ and $\hat Q$ between degenerate configurations in phonon coordinates, \emph{i.e.}, the distance between site 1 and site 2 and between the left ($l$) and right ($r$) wells in \autoref{fig:h_phi}.
%
Cubic spline interpolation is used to map the DFT potential energy onto the subgrid. 
Eigenvalues and wavefunctions are obtained using Implicitly Restarted Lanczos Method implemented in \texttt{scipy}; all results are converged with respect to the subgrid resolution. 

\begin{table*}
    \centering
    \caption{Computed tunnel splittings $J_\mathrm{H}$ and $J_\mathrm{D}$ (meV), zero-point energies ($ZPE$, meV), and parameters for each defect class by concentration ($c$). $R_H$ is the H–H site separation (\AA); $R_c$ the lattice displacement between local and symmetric configurations (\AA); $E_c$ and $E_c'$ the coincidence and effective coincidence energies (meV); and $V$ is the symmetric barrier height (meV).}
    \begin{ruledtabular}
    \begin{tabular}{lllcllcllll}
    Defect&$c$&  $R_\mathrm{H}$&$R_c$&$E_c$ &$E'_c$& $V$&$E^{ZPE}_\mathrm{H}$& $E^{ZPE}_\mathrm{D}$& $J_\mathrm{H}$ &$J_\mathrm{D}$\\
  \hline
 H& 1/16&  1.17&0.068& 17.9&19.9& 204.1& 240.4& 173.3&0.012& $4.9\times10^{-4}$\\
 & 1/54&  1.17&0.091& 16.8 &20.8& 200.3& 236.5& 170.0& $5.4\times10^{-3}$ & $2.7\times10^{-4}$\\
  \hline
  O-H&1/16&  1.05&0.044& 8.0&8.7& 152.4& 240.1& 173.3&0.31& 0.024\\
          &1/54&    1.10&0.069&12.5&13.5&  153.0&237.0& 170.3&0.064 &  $4.8\times10^{-3}$\\
 & 1/128& 1.08& 0.074&  12.8&-& -&  -& -&-& -\\
 \hline
 Ti-H& 1/16&  1.09&0.058& 11.2&12.3& 163.9& 232.1& 167.7&0.084 & $5.8\times10^{-3}$\\
 & 1/54&  1.11&0.078& 12.1 &13.2& 165.8& 228.4& 164.5&0.033 & $2.3\times10^{-3}$\\
 \hline
          Zr-H&1/16&    1.32&0.069&18.4&20.3&  279.9&236.3& 170.3& $1.5\times10^{-3}$ &  $2.5\times10^{-5}$\\
 & 1/16\footnote{Computed using 5D Hamiltonian.}& 1.32& 0.069& 18.4& 20.7\footnote{Evaluated using grid interpolation.}& 275.4\footnotemark[1]  & 245.3& -&$1.9\times10^{-3}$ &-\\
  &1/54&  1.31&0.087& 16.7&18.0& 273.6& 231.2& 166.1& $8.1\times10^{-4}$ & $1.5\times10^{-5}$\\
    \end{tabular}
    \end{ruledtabular}
    \label{tab:TunnelEnergy}
\end{table*}

\section{Results and Discussion}

\subsection{Localization and Lattice Coupling}
\autoref{fig:h_phi} shows 1D slices of the computed ground and first-excited state wavefunctions for an O-H defect along $\hat q_y$ for three values of the composite phonon coordinate $Q$. 
The order parameter $Q$ continuously transforms the Nb lattice between adjacent degenerate sites (upper left and lower right) through an intermediate symmetric configuration (upper right).
When $Q=Q_l$, the ground state wavefunction is almost entirely localized in the left well, consistent with the large well-to-well energy asymmetry ($54$ meV) between adjacent sites in the self-trapped state and prior work \cite{Sundell2004}. 
Here the Nb lattice locally deforms to accommodate the H atom localized in site 1 (\autoref{fig:h_phi}, upper left).
In the symmetric configuration  with $Q_s=Q_{lr}/2$, the wavefunction symmetrically spans both wells but with a reduced amplitude, whereas with $Q=Q_{lr}=Q_{r}$, localization of the wavefunction shifts to the right well 
(\autoref{fig:h_phi}, lower right).  
Correspondingly, the ground state wavefunction exhibits degeneracy breaking overlap when $Q\approx Q_s$, and highlights the strong coupling between the H motion and lattice distortion (\autoref{fig:h_phi}). %

The wavefunction of the first-excited state follows a similar trend; however, it undergoes a phase flip in the symmetric configuration that persists toward  the right well.
\autoref{fig:h_phi}g(h) further shows that the potential energy and wavefunction along the $Q$ coordinate has a harmonic character at each H minima, $\mathbf{q}_{1}$ ($\mathbf{q_2}$). We also plot the image of the potential energy surface at the other H minima, $\mathbf{q}_2$ ($\mathbf{q}_1$). Both surfaces meet at common coordinate, $Q_c$, and potential energy, $E_c'$, which defines the symmetric configuration and effective coincidence energy.

\subsection{Tunnel Splitting and Dimensionality Effects}
With the full 4D Hamiltonian, we compute a tunnel splitting ($J_\mathrm{H,D}$) ranging from 0.31 meV ($c=1/16$) to 0.064 meV ($c=1/54$) for hydrogen (H) and 0.024 meV  ($c=1/16$) to  $4.8\times10^{-3}$ meV ($c=1/54$) for deuterium (D) TLS in bcc Nb (\autoref{tab:TunnelEnergy}). 
$J_\mathrm{H}$ computed at $c=1/54$ is approximately 3-fold smaller than experimental estimates derived from specific heat (0.19 meV) and neutron scattering experiments (0.21 meV) \cite{Wipf1984}, consistent with later measurements reporting values ranging from 0.170 -- 0.230\,meV \cite{Magerl1986}. $J_\mathrm{D}$ computed at $c=1/54$ is about 4 times lower than the estimate from specific heat experiments (0.021 meV) \cite{Wipf1984p2}.
Interestingly, $J_H$ for self-trapped hydrogen is comparable to $J_D$ computed for oxygen-trapped hydrogen. This arises due to the confluence of a larger H--H site separation ($R_H$), a larger lattice displacement ($R_c$) and coincidence energy ($E_c$) between the local and symmetric configurations, and a larger barrier ($V$) in the symmetric configuration. 
To better understand the sensitivity of the tunnel splitting to the dimensionality of the Hamiltonian, we have also computed the tunnel splitting for lower-dimensional subspaces of the 4D Hamiltonian. The minimal subspace that produces a tunnel splitting is the 1D slice taken along $\hat q_y$ with $Q=Q_s$, yielding $J_\mathrm{H}=1.7$\,meV.
Increasing the dimensionality to account for the $q_x$ and $q_z$ modes reduces the computed $J_\mathrm{H}$ to
0.567 meV (\autoref{tab:Tunnel Dimension}). In the subspace spanning the $q_y$ and $Q$ modes, $J_H=0.360$\,meV. Correspondingly, the tunnel splitting is moderately sensitive to the anharmonic couplings between the H modes, which increase the curvature in the potential energy near the barrier maximum thereby suppressing wavefunction overlap. 
Although the defect concentration $c$ has minimal effects on these couplings, the tunnel splitting becomes more sensitive to $c$ when the $Q$ mode is included.

\begingroup
\begin{table}
    \caption{Effect of Hamiltonian dimension and concentration ($c$) on computed tunnel splitting ($J$, in meV).}
    \centering
    \begin{ruledtabular}
    \begin{tabular}{llllll}
 Defect& $c$& $J_\mathrm{4D}  {(\mathbf{q},Q})$& $J_\mathrm{2D} (q_y,Q)$& $J_\mathrm{3D} (\mathbf{q})$&$J_\mathrm{1D} (q_y)$\\
 \hline
 H& 1/16& 0.012& 0.079 &0.17& 0.53\\
 & 1/54& $5.4 \times 10^{-3}$& 0.046 &0.18 & 0.57\\[0.3em]
  O-H&1/16& 0.31& 1.0 &0.92 & 2.2\\
          &1/54&   0.064& 0.360 &0.57 & 1.7\\[0.3em]
 Ti-H& 1/16& 0.084& 0.46 &0.49 & 1.4\\
 & 1/54& 0.033& 0.24 &0.40 & 1.2\\[0.3em]
          Zr-H&1/16&   $1.5 \times 10^{-3}$& $7.1\times10^{-3}$&0.021 & 0.059\\
  &1/54& $8.1\times 10^{-4}$& $5.2\times10^{-3}$&0.021& 0.067\\
  \hline
 D& 1/16& $4.9\times10^{-4}$&  $2.7\times10^{-3}$&$6.5 \times 10^{-3}$& 0.021\\
 & 1/54& $2.7\times10^{-4}$&  $1.6\times10^{-3}$&$7.6 \times 10^{-3}$& 0.024\\[0.3em]
 O-D& 1/16& 0.024&  0.077&0.073& 0.18\\
 & 1/54& $4.8\times10^{-3}$&  0.023&0.040& 0.12\\[0.3em]
 Ti-D& 1/16& $5.8\times10^{-3}$&  0.027&0.031& 0.094\\
 & 1/54& $2.3\times10^{-3}$&  0.013&0.024& 0.073\\[0.3em]
 Zr-D& 1/16& $2.5 \times 10^{-5}$&  $9.2\times10^{-5}$&$3.1\times10^{-4}$& $8.4\times10^{-4}$\\
 & 1/54& $1.5 \times 10^{-5}$&  $7.5\times10^{-5}$&$3.2\times10^{-4}$& $9.1\times10^{-4}$\\
    \end{tabular}
    \end{ruledtabular}

    \label{tab:Tunnel Dimension}
\end{table}
\endgroup

Given the sensitivity of the tunnel splitting to the anharmonic terms in the Hamiltonian, we have further validated our results by comparing the computed transition frequencies between the ground state and  excited states against measured inelastic neutron scattering data for O-H and H defects \cite{Magerl1983}. The reported transition frequencies are 108 (106) meV and 159 (163) meV for O-trapped H (self-trapped H), with the latter transition two-fold degenerate due to the local symmetry of tetrahedral sites.
Our results reproduce these values within a few meV at both defect concentrations (\autoref{tab:excitedstates}).
By including the composite phonon mode $Q$, we find additional transitions appear in the excitation spectrum of the 4D Hamiltonian with frequencies dictated by the nearly harmonic states of the $Q$ mode. As the energy of these excited states overlap with the excited-state energies of the H-dominant modes, assessing the appropriate transition frequencies required a manually assignment. This was performed by plotting the wavefunctions of the excited states and selecting the excited states whose nominal wavefunction was consistent with the harmonic oscillator wavefunction of the desired transition.
The transition energy between the ground and first excited state varied between 103-110 meV for all evaluated defects, indicating a weak sensitivity to barrier height. 
The transition energy between the ground state and higher quasi-degenerate excited states ranged between 140--173\,meV depending on $c$ and defect type. 
These results confirm that our sampled potential $V(\mathbf{q},Q)$ accurately captures the relevant O-H energy landscape and represent a clear improvement over the prior 3D Hamiltonian approach \cite{Sundell2004}, likely due to the higher fidelity DFT functional choice and denser sampling.

\begin{table}
   \caption{Computed excited state transition frequencies and tunnel splittings for  defects (in meV) with concentration ($c$).}
    \centering  
    \begin{ruledtabular}
    \begin{tabular}{llcccccc}
          Defect&   $c$&  $\hbar \omega_1$& $\hbar \omega_2$&$\hbar \omega_3$ & $J_1$& $J_2$&$J_3$\\
          \hline
 H& 1/16& 110& 153& 156& 0.17& $3.3 \times 10^{-2}$&0.34\\
 H& 1/54& 108& 151& 155& 0.071& $9.7 \times 10^{-5}$&0.21\\
 H\footnote{Values derived using neutron spectroscopy at 295\,K in \cite{Magerl1983}.}& 1/180& 106& 163& 163& -& -&-\\
 \hline
  O-H&  1/16& 106& 160& 173& 4.7& 5.3&$2.6\times10^{-2}$\\
  O-H&  1/54& 106& 149& 155& 0.14& 1.9&$2.4\times10^{-3}$\\
 O-H\footnote{Values derived using neutron spectroscopy at 4\,K in \cite{Magerl1983}.}& 1/100& 108& 159&159& -& -&-\\
  \hline
 Ti-H& 1/16& 104& 151& 160& 1.26& $3.1\times10^{-2}$&0.14\\
 Ti-H& 1/54& 103& 143& 148& 0.53& 0.22&$7.9\times10^{-4}$\\
 \hline
 Zr-H& 1/16& 109& 144& 167& 0.023& $8.5\times10^{-5}$&0.094\\
 Zr-H& 1/54& 106& 140& 164& 0.011& $2.1\times10^{-5}$&0.052\\
    \end{tabular}
    \end{ruledtabular}
     \label{tab:excitedstates}
\end{table}

\subsection{Impact of Nb Mode Coupling}
The effect of the harmonic Nb mode couplings on $J$ is expected to be small. 
At the symmetric configuration with a potential energy minima at $Q=Q_{s}$,  the wells are degenerate, however, second-order mode couplings $\omega_{q_iq_Q}$ to the phonon bath slightly offset and lower the local minimum from $\{q_i\}_{\ne Q}=0$. 
Degenerate wells remain following this offset, which reduces the transition energy and 
marginally increases the tunnel splitting. 
We estimate the maximum reduction by comparing the energy difference between the effective, $E_c'$, and true, $E_c$, coincidence energies when all other modes are fully relaxed at $Q=Q_s$. 
This energy difference is typically between $1-2$ meV lower than the symmetric structure, corresponding to about $10-20\%$ of the total symmetrization energy.  For the O-H defect, $E_c$ is only 1 meV lower than $E_c'$; such a minor correction cannot reconcile our computed $J=0.064$\,meV ($c=1/54$) with the experimental value of $J_\mathrm{H}=0.19$\,meV. Indeed, this result follows from the fact that $E_c'$ for $c=1/16$ is 5 meV lower than $E_c'$ for $c=1/54$, yet $J_\mathrm{H}$ is only 50\% larger than the experimental 
value.

Anharmonic couplings between Nb lattice modes and H modes near the barrier maximum provide another  mechanism to increase $J$.
A prior work has shown that the barrier height along the MEP is significantly lowered compared to the symmetric configuration \cite{Sundell2004}.
At the symmetric configuration, $Q=Q_s$, the potential is approximately quadratic in each well as 
determined by the phonon structure with small offsets determined from the coupling $\omega_{qQ}$. 
%
Near the potential maximum, however, the  
MEP constraint requires that a finite population of Nb modes lowers the effective barrier height, 
correspondingly, these modes enhance wavefunction overlap and introduce a first-order correction that increases $J$.
Therefore, our computed tunnel splitting is a conservative lower bound to the actual tunnel energy. 
Since the H coupling to the Nb modes reduces the barrier height, lower-dimensional calculations, \emph{i.e.}, where lattice distortions are partially excluded, cannot serve as strict upper bounds.

\begin{figure}
\centering
   \includegraphics[width=0.49\textwidth]{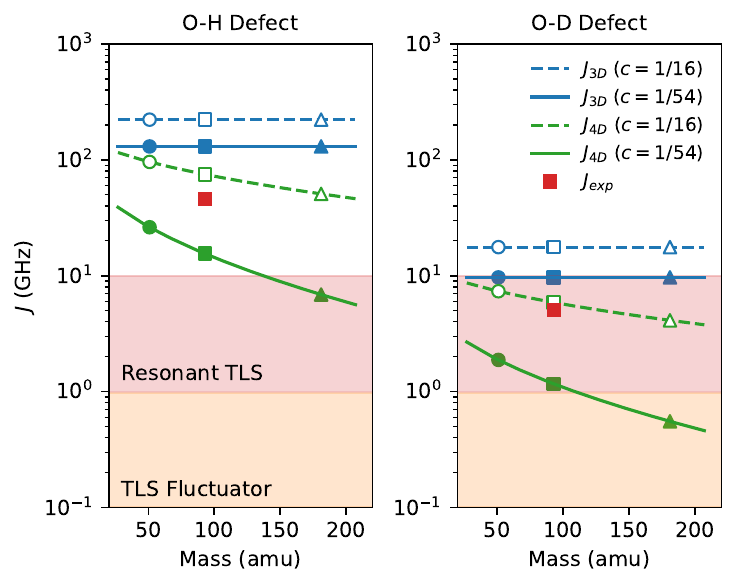}
   \caption{Exponential dependence of the computed tunnel splitting ($J$) for O-H and O-D defects on the mass-scaled phonon coordinate $Q'=Q\sqrt{m/m_\mathrm{Nb}}$ using a 4D Hamiltonian. Markers indicate  $m=m_\mathrm{V}$ (circles), $m=m_\mathrm{Nb}$ (squares), and $m=m_\mathrm{Ta}$ (triangles). Experimental (exp) data from \cite{Wipf1984p2}.} 
   \label{fig:ec_mode}
\end{figure}

\subsection{Mass Effects}
The dominant anharmonicity arises from the cross-coupling $\mathbf{q}-Q$ between the H modes and the composite Nb lattice mode. 
This interaction strongly influences tunneling behavior and challenges the validity of the light-particle picture.
For a harmonic oscillator, the wavefunction amplitude $\psi\propto \exp{(-\omega Q^2/(2\hbar))}$.
The mode frequency along the composite coordinate is given by $\omega=\sqrt{2E_c}/Q_c$, which leads to the ratio:
%
$\psi(Q_c)/\psi(0)= \exp\left(-E_c/\hbar\omega\right)$. 
By scaling $Q$ over a wide range of values, we evaluate how the mode frequency and coincidence energy affect the tunneling frequency of O-H and O-D defects. This is readily achieved by replacing $Q$ with $\sqrt{m/m_\mathrm{Nb}}Q$.
Physically, this rescaling either enhances ($m<m_\mathrm{Nb}$) or suppresses ($m>m_\mathrm{Nb}$) tunneling by increasing (decreasing) the magnitude of the wavefunction in the symmetric configuration ($Q=Q_s$) and thereby increasing (decreasing) wavefunction overlap.
\autoref{fig:ec_mode} shows that tunnel splitting decreases by approximately a factor of 2-3 when moving from Nb to Ta. 
We also find that $J=J_\mathrm{4D}$ can vary by nearly an order of magnitude compared to the light-particle value ($J_\mathrm{3D})$, highlighting the critical role of these anharmonic couplings in renormalizing tunnel splittings.
This reduction shifts the expected lower-bound resonant frequency of H-based TLS into the 1--20\,GHz range. However, strain coupling energies of 0.2--1 eV imply that even small lattice strains (as low 0.002-0.02\%) are sufficient to raise the resonant frequency above typical superconducting qubit operating  frequencies, effectively suppressing most TLS from acting as resonant defects.

\subsection{Acoustic Phonons and Strain Interactions}
Configurational TLS couple to lattice strain, which modulates the energy asymmetry between wells \cite{Phillips1981}. 
This coupling is dominated by low-energy acoustic modes of the host crystal. 
Our formalism reveals an additional broadband interaction: the TLS ground state couples to lattice phonons through the composite coordinate $Q$.
This degree of freedom introduces new channels for energy exchange beyond conventional strain coupling. 
In the ground state, strain coupling is strongly suppressed: tunneling defects with no intrinsic well-to-well asymmetry exhibit only second-order strain interactions. 
Consequently, strain is required to break the degeneracy in the stress-free state. 
In practice, equilibrium configurations are rarely achieved in thin films due to finite-size effects, making it essential to evaluate strain effects across a range of hydrostatic and deviatoric strains relative to the infinite-lattice geometry.
In the harmonic limit, the nuclear Hamiltonian near a minimum may be expanded as
$    E^{(i)}/V_0 = C_{jklm}^{(i)}\varepsilon_{jk}\varepsilon_{lm}/2$,  
where $C_{jklm}^{(i)}$ is the elastic stiffness tensor and $\varepsilon_{jk}$ the strain components. 
Under a non-zero macroscopic strain field, $\alpha^{(i)}_{jk}$, the energy of site $i$ becomes
$    E^{(i)}/V_0 = \frac{1}{2}C_{jklm}^{(i)}(\varepsilon_{jk}-\alpha^{(i)}_{jk})(\varepsilon_{lm}-\alpha^{(i)}_{lm})$, which contains the first-order term:
\begin{equation*}
    \frac{E^{(i)}}{V_0} = -\frac{C_{jklm}^{(i)}}{2}\left(\varepsilon_{lm}\alpha_{jk}^{(i)}+\varepsilon_{jk}\alpha_{lm}^{(i)}\right)=-C_{jklm}^{(i)}\varepsilon_{lm}\alpha_{jk}^{(i)}\,,
\end{equation*}
using the symmetry $C_{jklm}=C_{lmjk}$. The relative energy between two two adjacent sites is then
$\Delta E^{(ij)}/{V_0} = (\sigma^{(i)}_{lm}-\sigma^{(j)}_{lm})\varepsilon_{lm}\,,$
where $\sigma^{(i)}_{lm}=C_{jklm}^{(i)}\alpha^{(i)}_{jk}$.
%
%
After multiplying by the domain volume, we obtain 
$\Delta E^{(ij)}=(P^{(j)}_{lm}-P^{(i)}_{lm})\varepsilon_{lm}\,,$ where
$P^{(i)}_{lm}$ is  the elastic dipole tensor of the defect.
Thus, in the dilute limit, site-to-site asymmetry of a given tunneling system depends depends only on the local stress field at the defect, consistent with the elastic dipole tensor formalism \cite{Wrbel2021}.

This framework demonstrates that local strain fields can strongly modulate TLS energetics, making strain engineering and control over homogeneity critical for mitigating decoherence \cite{Chen2024};
highly pure superconducting films with low internal strain fields, will be the most prone to resonant TLS and multilevel systems (MLS).
In crystalline systems, only TLS subject to nearly isotropic or likewise symmetry-preserving stress/strain fields will exhibit tunneling behavior, as tunneling is effectively quenched above a critical well-to-well asymmetry.

\begin{table}
    \centering
        \caption{Computed eigenvalues, $E_i$ (meV), for Zr-H four-level systems with $c=1/16$. Eigenvalues are reported relative to $(E_1+E_2)/2$ to facilitate comparison with prior work \cite{Cannelli1994}.}
    \begin{ruledtabular}    \begin{tabular}{lcccc}
 Defect& $E_0$&$E_1$& $E_2$& $E_3$\\
 \hline
         Zr-H&  $-1.9\times 10^{-3}$& $-1.2 \times 10^{-5}$& $1.2 \times 10^{-5}$& $1.9\times 10^{-3}$\\
    \end{tabular}
    \label{tab:placeholder}
    \end{ruledtabular}
    \label{tab:efls}
\end{table}

\subsection{Four-Level and Multi-Level Systems}
Substitutional defects in bcc Nb preserve its cubic point group symmetry.
%
Adjacent to any substitutional impurity there are 24 degenerate tetrahedral sites.
This high-site degeneracy leads to complex energy-level structures for isolated H defects which occupy those sites. 
The H site degeneracy, however, is affected by internal stress fields that 
reduce the number of degenerate or nearly-degenerate H sites \cite{Cannelli1994, Cordero2022}.
Indeed, anelastic-relaxation experiments on NbZr$_{0.0045}$H$_x$  only supports the existence of a population of H four-level systems (FLS) \cite{Cannelli1994, Cordero2022}. 
To model FLS, we extend our formalism by introducing two quasi-orthogonal lattice coordinates $S$ and $T$. 
Each coordinate $S$ or $T$ transforms singly-degenerate lattice configurations into a four-fold degenerate state (\autoref{fig:h_phi}, bottom left). 
Their combined action, $Q=\left(S-T\right)/\sqrt2$, recovers the order parameter used to model TLS but now spans a higher 5D configurational space. 
This extension enables us to capture the coupling between multiple wells. 
We now evaluate the H-level structure for a Zr substitutional defect with $c=1/16$ (\autoref{tab:efls}). From \cite{Cannelli1994}, the level structure computed using a four basis element Hamiltonian with a nearest-neighbor coupling term ($J/2$) and degenerate wells will be $-J$, 0, 0, $J$, where the eigenvalues are referenced to the nominal zero-point energy of each well.
The computed tunnel splitting using our FLS model ($J=-E_0$) is approximately $1.9\,\mu$eV, compared to $1.5\,\mu$eV for the corresponding TLS model, while $E_c^\prime$ remains essentially unchanged from the Zr-H TLS configuration. In addition, the third excited state $E_3$ is equivalent to $-E_0$ indicating that next-nearest neighbor coupling is negligible.
These results validate the nearest-neighbor coupling model proposed in \cite{Cannelli1994} and show that the inclusion of an additional phonon mode does indeed increase $J$ and that the enhancement of the tunnel splitting is captured due to the distortion of the tunneling path along the new lattice mode and not through additional tunneling through the four-fold degenerate structure.
In the specific case of a four-site Hamiltonian, the level structure under tetragonal strains possess both hyperbolic and linear dependencies on applied strain \cite{Cordero2022}. 
Consequently, FLS are more likely to maintain transition energies near qubit operating frequencies under realistic internal strain fields present within typical films, making them a potential source of enhanced decoherence in superconducting devices.
In reality, substitutional defects in Nb generate MLS described by an effective 24-dimensional Hamiltonian. The exact level structure will depend on the relative well energies which in turn depends on the local-stress state (as shown above) and short-range defect-defect interactions. Like FLS, these MLS will possess a level structure essentially bounded by the hyperbolic transition-energy curves of a given TLS subspace, leaving intermediate transitions within a fixed upper limit; the lower limit is bound by zero. 
Consequently, even substantial lattice strains cannot fully eliminate such MLS 
from contributing as resonant or subresonant states under stress fields that would typically quench two-level behavior. 
These MLS maintain transition energies within the qubit frequency range over a broader span of internal strain fields; however, due to the increased site degeneracy, this behavior persists only when the local strain tensor is nearly isotropic. For example, Zr-H defects in the dilute limit have been experimentally shown to support at most a four-level structure, far below the theoretical upper bound of 24 \cite{Cannelli1994, Cordero2022}.
While heavier lattice masses and strain fields suppress TLS activity, MLS remain robust over a wider range of conditions, making them a critical consideration for understanding decoherence in superconducting qubits.

\subsection{Tunnel splittings from machine-learned potentials}

A recent work employed a machine-learned potential (MLP) to compute tunnel splittings for  O-trapped H in bcc Nb, reporting $J=0.275$\,meV for the low-energy trapping site corresponding to our site 7a, and $J=0.414$\,meV for a higher energy site (equivalent to our site 5a). 
These values were obtained using a static lattice in a symmetric configuration following a procedure closely related to that described in Sec.~\ref{appendix:structures}. 
%
We note that the identification of the most stable O-H configuration and the associated relaxation protocol match those employed, although not documented, in an earlier draft of the present work \cite{Pritchard2025b}, which predates the results of Ref.~\cite{Abogoda2025}.

The $J$ reported in Ref.~\cite{Abogoda2025} is most directly comparable to our static-lattice DFT result in \autoref{tab:Tunnel Dimension}, where we obtain $J_{3D}=0.57$\,meV for for O-trapped H at site 7a. 
The MLP used in Ref.~\cite{Abogoda2025} is reported to achieve a mean absolute error in the computed total energy per atom of less than 1 meV/atom, which is below the typical $\sim$5 meV/atom threshold typically considered acceptable for high-quality MLPs. 
However, no errors in the predicted atomic positions were reported, despite the fact that even picometer deviations in H coordinates and lattice positions can strongly impact the computed tunnel splitting.

This comparison emphasizes an important limitation in interpreting tunnel splittings obtained from MLPs: Even for chemically simple systems, such as O-trapped H in bcc Nb, uncertainties intrinsic to the MLP parametrization introduce an additional source of error beyond those inherent to the underlying DFT functional. 
%
While a well-trained MLP may produce splittings in reasonable agreement with experiment, such agreement alone does not validate either the parametrization strategy or the theoretical formalism and may in fact be accidental. 
Apparent improvements over DFT can arise from error cancellation within the interatomic potential and should be acknowledged. 
Similar forms of cancellation, and hence accidental agreement, are documented and occur in high-throughput low-fidelity DFT studies \cite{Sohier_2021,Ha2024}.
%
%

Our results show that a static-lattice treatment at the DFT level systematically overestimates the experimental tunnel splitting, yielding $J_\mathrm{3D}$ values larger than measured. 
In contrast, inclusion of coupling to the composite mode $Q$ suppresses the splitting substantially, producing 
$J_\mathrm{4D}$ values which bound the experimental $J$ from below by an equal margin.
We show in a forthcoming work that the inclusion of the lattice mode which passes through the transition state alleviates most of the remaining error in the computation of $J$.
%
We would expect the same qualitative behavior, \emph{i.e.}, an over-estimation of $J$ in the static‑lattice approximation and a large reduction when coupling to the $Q$ mode is included, using the MLP of Ref.~\cite{Abogoda2025}.

\section{Conclusions}
Standard approaches in the superconducting qubit community fail to capture essential features of two-level and multilevel defects arising from degenerate minima in the nuclear Hamiltonian. Symmetrized structures significantly overestimate tunnel splittings and are not thermodynamically stable, particularly for configurations with large coincidence energies. We address this limitation by introducing composite phonon coordinates that account for lattice displacements between degenerate wells, enabling a lattice-renormalized formalism for configurational tunneling systems.
Our calculations for O-H and O-D tunneling defects in bcc Nb demonstrate that a 4D Hamiltonian provides a lower bound on experimental tunnel splittings 0.064\,meV (0.0048\,meV) compared to 0.19\,meV (0.021\,meV) for oxygen-trapped H (D) TLS in bcc Nb, while a 3D model without $Q$ bounds them from above 0.57\,meV (0.040\,meV). 
Additionally, $J_\mathrm{H}$ computed for self-trapped hydrogen is comparable to $J_\mathrm{D}$ computed for oxygen-trapped hydrogen. This finding supports the conclusion that almost all H is either trapped by defect sites or precipitates to the $\varepsilon$-phase at mK temperature; as otherwise, the measured specific heat anomaly should extend below that observed in \cite{Wipf1984, Wipf1984p2}.
Extending to a 5D Hamiltonian for a FLS enhances the splitting by 27\%, primarily due to bending of the tunneling path enabled by additional phonon coupling. These results highlight the critical role of anharmonic coupling to lattice phonons near the barrier maximum, suggesting that further refinement of the model will improve agreement with experiment.

By explicitly computing the formation energy of H tetrahedral sites adjacent to each trapping defect, we show that O, Ti, and Zr-trapped H exhibit  unambiguously favored sites.
Each of the thermodynamically favored sites possess at least one adjacent, degenerate site which validates the assumption of active tunneling systems in H-doped Nb with secondary trapping defects \cite{Neumaier1982, Wipf1984, Cannelli1994}.
In particular, O-trapped H hosts only 2 adjacent, degenerate sites supporting the two-level model proposed in \cite{Wipf1984}.
In contrast, Ti- and Zr-trapped H possess 24 adjacent, degenerate sites that  support  possible multi-level system behavior of these defects \cite{Cannelli1994, Cordero2022}.
The preference for these sites are well-explained by simple elastic and quantum mechanical bonding arguments.

Our formalism extends naturally to multilevel systems, which remain active under strain conditions that suppress TLS-qubit interactions and thus represent an important source of decoherence in superconducting qubits. 
By linking tunneling dynamics to phonon-mediated strain interactions, we show that local stress fields, via elastic dipole coupling, strongly modulate defect energetics while the composite coordinate $Q$ introduces additional channels for energy exchange beyond conventional strain coupling.
We further note that the TLS density for amorphous materials is on the order of $0.1$\,eV$^{-1}$\,nm$^{-3}$ \cite{Berret1988}.
The H TLS density may be estimated as $\rho_\mathrm{H}Z(E)\approx2\rho_H/\pi\varepsilon_0$ \cite{Wipf1984}. Using $\varepsilon_0=4.5$\,meV and $\rho_\mathrm{H}=0.43$ atoms\,nm$^{-3}$ ($c=1/128$), we find the TLS density to be approximately 60\,eV$^{-1}$\,nm$^{-3}$. Correspondingly, even a dilute density of interstitial H introduces several orders of magnitude more TLS than the same volume of a glassy material.
These insights underscore the need for strain engineering and homogeneity control in superconducting films and provide a predictive framework for designing materials that minimize TLS-related loss in superconducting quantum circuits.

\begin{acknowledgments}
This work was supported by the U.S. Department of Energy, Office of Science, National Quantum Information Science Research Centers, Superconducting Quantum Materials and Systems Center (SQMS), under Contract No.\  89243024CSC000002. Fermilab is operated by Fermi Forward Discovery Group, LLC under Contract No.\ 89243024CSC000002 with the U.S.\ Department of Energy, Office of Science, Office of High Energy Physics.
This research used resources of the National Energy Research Scientific Computing Center, a DOE Office of Science User Facility supported by the Office of Science of the U.S.\ Department of Energy under Contract No.\ DE-AC02-05CH11231 using NERSC award BES-ERCAP0023827.
This research was supported in part through the computational resources and staff contributions provided for the Quest high performance computing facility at Northwestern University which is jointly supported by the Office of the Provost, the Office for Research, and Northwestern University Information Technology. 
We thank Dominic Goronzy and David Garcia-Wetten for helpful discussions on the nature of hydrogen incorporation in bcc Nb.
\end{acknowledgments}

\section*{DATA AVAILABILITY}
The data that support the findings of this article are available in \cite{supp} and online at 
\cite{supp_dryad}.\\

\section*{Appendix}
\appendix
\renewcommand{\thefigure}{A\arabic{figure}} %
\setcounter{figure}{0} 
\renewcommand{\thetable}{A\Roman{table}} 
\setcounter{table}{0} 

\section{Derivation of the reduced Hamiltonian\label{appendix:derivation}}

We begin from the Born–Oppenheimer nuclear Hamiltonian
\begin{equation}\label{eq:bo_appendix}
    \hat H_n=\sum_i\frac{-\hbar^2}{2m_i}\nabla_i^2 + V(\{\vec r_i\}),
\end{equation}
where $m_i$ are nuclear masses for atom $i$, $\nabla_i^2$ are Laplacians, $V$ is the potential energy, and $\{\vec r_i\}$ are nuclear coordinates referenced to a chosen configuration. 
Here it is convenient to choose the center-of-mass frame of the lattice \textit{without} a H interstitial. One may then, in principle, separate $V$ into its harmonic and anharmonic components; however, a natural issues arises: for tunneling systems with multiple degenerate or nearly degenerate minima, a single harmonic expansion is inadequate. We therefore partition configuration space into Voronoi cells $\{\Omega_k\}$ centered at each local minimum $k$ of the hydrogen defect and introduce the  function $H_k(\vec r_H)$ such that $H_k=1$ for $\vec r_H\in\Omega_k$ and $H_k=0$ otherwise, i.e., defined in a similar fashion to the Heavyside step function. 
$H_k(\vec r_H)$ ensures local harmonicity while allowing multiple basins. This construction avoids imposing a globally symmetrized (thermodynamically unstable) structure and is standard in multi-well problems.

Expanding the potential locally in each basin $k$ about the equilibrium positions $\{\vec r^{\,eq,k}_i\}$ yields
\begin{widetext} 
\begin{align}\label{eq:piecewise_harm}
    V(\{\vec r_i\}) = \sum_{k} H_k(\vec r_H)\left[
    V_k^{(0)} + \frac{1}{2}\sum_{i j}\sum_{\mu\nu}
    \left(r_{i,\mu}-r_{i,\mu}^{\,eq,k}\right)
    \Phi^{(k)}_{i\mu,j\nu}
    \left(r_{j,\nu}-r_{j,\nu}^{\,eq,k}\right)
    \right] + V_{\text{anharm}}(\{\vec r_i\}),
\end{align}
\end{widetext}
where $\Phi^{(k)}$ is the Hessian in basin $k$, and $V_{\text{anharm}}$ collects higher-order and inter-basin couplings.
(Here $i$ and $j$ are atom indices, $k$ indexes each degenerate H site, and $\mu$ and $\nu$ are the cartesian indices for the atomic coordinates $\vec r_i$ and $\vec r_j$.)
%

The mass dependence of \autoref{eq:bo_appendix} may be eliminated by using the standard transformation to phonon coordinates:
\begin{equation}
q_i=C_{ij}^\dagger m_j^{1/2}\delta_{jk}r_k
\end{equation}
Where, $C_{ij}$ is the matrix of phonon eigenvectors chosen with respect to a stable nuclear configuration, $m_j$ are the nuclear masses, and $\delta_{jk}$ is the Kronecker delta. After applying this transformation the nuclear Hamiltonian may be written as:
\begin{equation}
\hat H_n=\sum_i\frac{-\hbar^2}{2}{\nabla'}_{i}^{2} +\sum_{ik}\frac{1}{2}\omega_{ik}^2 q_{ik}^2H_k(\vec q_\mathrm{H})  +V_{anharm}(\{q_i\})
\end{equation}
Where $\nabla'$ indicates the mass-scaled $\nabla$ operator, $w_{ij}$ and $q_{ij}$ represent the frequency and mode amplitude of phonon mode $i$ in the harmonic well $k$.  For clarity, we do no explicitly distinguish the longitudinal and transverse modes. We also make the approximation that $\mathbf{q}=\vec q_\mathrm{H}=m_\mathrm{H}^{1/2} \vec r_\mathrm{H}$ which is justified as the phonon eigenvector of the H modes in bcc Nb posses essentially no contribution from Nb atoms.  
Because the kinetic energy is invariant under orthonormal changes of basis in mass-weighted coordinates, the Laplacian transforms as $\sum_i(-\hbar^2/2m_i)\nabla_i^2 \to \sum_{\alpha}(-\hbar^2/2)\,\partial^2/\partial Q_{\alpha}^2$. This justifies the use of the mass-scaled operator $\nabla'$.

We posit that $V_{anharm}$ is dominated by the anharmonicity due to the H phonon modes and the composite phonon coordinate Q which transforms the Nb lattice between two degenerate wells (\autoref{fig:h_phi}):
\begin{equation}
\hat H_n\approx\sum_i\frac{-\hbar^2}{2}{\nabla'}_{i}^{2} +\sum_{ik}\frac{1}{2}\omega_{ik}^2 q_{ik}^2H_k(r_H)  +V({\vec q_H,Q})
\end{equation}
\begin{equation}
 \vec Q=M^{1/2}\vec R=m_i^{1/2}\delta_{ij}R_j
\end{equation}
where, $\vec R$ is the displacement vector which transforms the non-Hydrogen atoms between their adjacent equilibrium sites and $Q$ is the Euclidian norm of $\vec Q$. We are always free to rotate the phonon Hamiltonian such that $Q$ is a basis vector. If we simultaneous re-diagonalize the phonons in the subspace without the $Q$ mode and the H modes we have:
\begin{widetext}
\begin{align}
\hat H_n \approx \sum_{i\ne Q,H}\frac{-\hbar^2}{2}{\nabla'}_{i}^{2} + \sum_{i\ne Q,k}\frac{1}{2}\omega_{ik}^2 q_{ik}^2H_k(\vec q_H)+  \sum_{i\ne Q,k}\frac{1}{2}\omega _{iQk}^2q_{ik}q_{Qk}H_k(\vec q_H)+ \sum_{i= Q,H}\frac{-\hbar^2}{2}{\nabla'}_{i}^{2}+V({q_H,Q})
\end{align}
\end{widetext} 
where we have incorporated the harmonic components of the $Q$ and H modes into $V$. This partially recouples the potential energy in phonon coordinates but allows us to separate the Hamiltonian into a 4 dimensional subspace which may be evaluated independently if we ignore the cross-terms between the composite phonon coordinate Q and the remaining phonon modes, giving Eq.~\ref{eq:reduced}.

The strength of the cross-couplings may be understood by realizing that that along a composite phonon coordinate $\hat Q'=\sum_{k} a_k \hat q_k$, tangent modes may lower the total energy unless $\hat Q'$ is a an eigenmode. In our model, we define $\hat Q$ based on the relaxed lattice configuration. Due to the small displacements required to achieve this transformation, the effective coincidence energy, $E_c'$, along this path is solely determined by phonon frequencies. The true coincidence energy, $E_c$, which is based on the full structural relaxation in the degenerate configuration includes the effect of additional phonon coordinates. We find that $E_c\approx0.8-0.9E_c'$ which implies the existence of one dominant phonon mode with additional broadband couplings mediating the transformation. Therefore, cross-coupling will tend to be weak for $E_c\approx E_c'$.

\section{Nb Phonons}
To evaluate the vibrational coupling between Nb, O, and H ions in bcc Nb with an O-H defect, we  computed the $\Gamma$-point phonon frequencies and eigenvectors for the $c=1/128$ structure with an O-H defect using the Phonopy code \cite{phonopy-phono3py-JPSJ}. 
The computed H mode frequencies are 29.5, 41.4, and 41.5\,THz, whereas the O mode frequencies are 9.97, 11.4, and 19.0\,THz. 
The phonon eigenvectors were converted to relative displacements by scaling them by the mass matrix $\mathbf{M}^{-1/2}=m_i^{-1/2}\delta_{ij}$, where $m_i$ is the mass of ion $i$. The computed H displacement component of each H mode exceeded 0.9999, while for each O mode 
the component of the displacement exceeded 0.99. 
These results strongly justify our treatment of the H modes as fully decoupled from all other atoms.

\begin{figure}
\centering
   \includegraphics[width=0.52\textwidth]{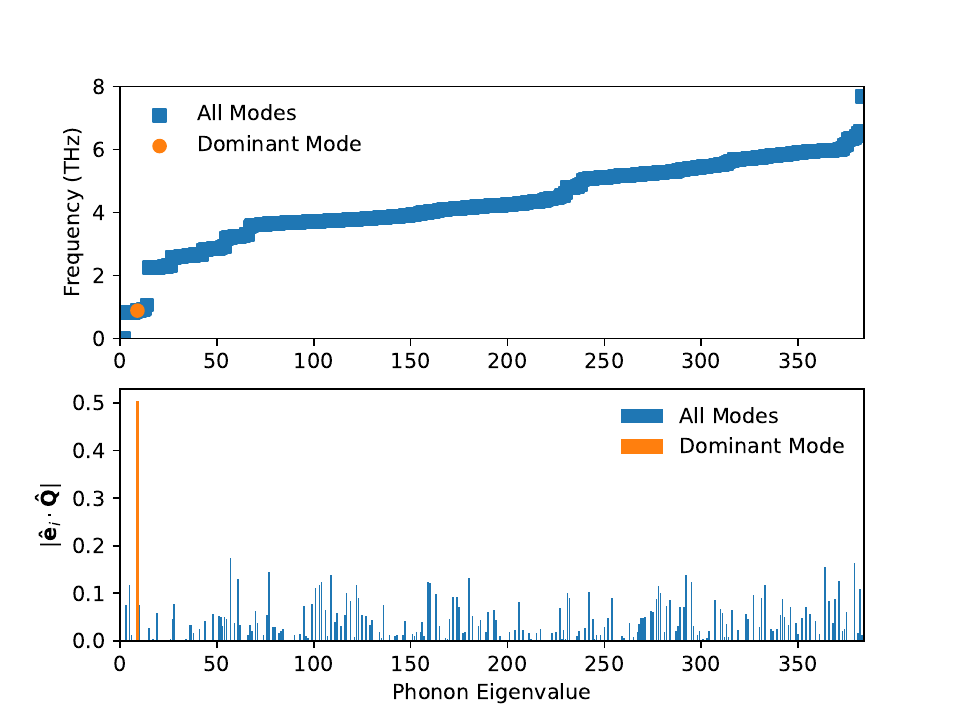}
   \caption{(top) $\Gamma$-point phonon frequencies for the OH defect structure ($c=1/128$) in bcc Nb. (bottom) Projection of the composite coordinate unit vector $
   \mathbf{\hat Q}$ onto each phonon eigenvector $\mathbf{\hat e}_i$. Oxygen and hydrogen modes are excluded in both panels.} 
   \label{fig:mode_contributions}
\end{figure}

\autoref{fig:mode_contributions} shows the distribution of mode frequencies and the projection of the composite mode $\mathbf{\hat Q}$ onto the phonon eigenvectors $\mathbf{\hat e}_i$. A single dominant, low-frequency mode is present with a projected value of 0.50. The remaining modes with non-zero projections are broadly distributed across the phonon frequency spectrum---consistent with the localized displacements which transform the lattice between degenerate configurations.

\section{Coincidence Structures\label{appendix:structures}}
In Ref.\ \cite{Sugimoto1980}, a theory is introduced in which light tunneling particles are treated  adiabatically with respect to lattice. The lattice relaxes under an effective force that can be derived from the Hellman-Feynman theorem:
\begin{equation}
\label{eq:nonlocal_force}
\mathbf{F_n}=\int d\mathbf{r}\left|\psi(\mathbf{r})\right|^2\mathbf{f}_n(\mathbf{r})\,,
\end{equation}
where $\psi(\mathbf{r})$ is the wavefunction of the light particle, $\mathbf{f}_n(\mathbf{r})$ is the force on lattice ion $n$ due to the light particle located at position $\mathbf{r}$, and $\mathbf{F}_n$ is the resulting effective force on ion $n$. This formulation is conveniently adapted for computing coincidence structures. 

When the probability density is approximated as a  superposition of two delta functions centered at the most probable positions of the light particle, the effective force on the lattice reduces to:
\begin{equation}
\mathbf{F}_n=\left|\psi(\mathbf{r}_1)\right|^2\mathbf{f}_n(\mathbf{r}_1) + \left|\psi(\mathbf{r}_2)\right|^2\mathbf{f}_n(\mathbf{r}_2)\,.
\end{equation}
An equivalent equation can be derived for the stress tensor $\sigma_{ij}$. For H or D  interstitials in bcc Nb, the coincidence configuration corresponds to equal wavefunction population in each potential well. We emphasize that relaxing the lattice using \autoref{eq:nonlocal_force} is equivalent to a constrained relaxation of the lattice subject to an equal population constraint. \autoref{eq:nonlocal_force} was implemented in combination with the BFGS optimizer in the Atomic Simulation Environment \cite{HjorthLarsen2017} (ASE) to obtain fully relaxed coincidence structures. The coincidence energy, $E_c$, is defined as the relative energy difference between the coincidence structure and the self-trapped configuration. As shown in \autoref{tab:TunnelEnergy}, the coincidence energy determined obtained using \autoref{eq:nonlocal_force} is approximately 10\% lower than the effective coincidence energy $E_c'$.

\bibliography{bib}

\end{document}